\documentstyle[12pt]{article}
\def\a{\alpha}
\def\d{\delta}
\def\b{\beta}

\def\beq{\begin{equation}}
\def\eeq{\end{equation}}
\def\gord{$ \raisebox{-.3ex}{$\stackrel{>}{_{\sim}}$} $}
\def\lord{$ \raisebox{-.3ex}{$\stackrel{<}{_{\sim}}$} $}

\begin{document}
\begin{titlepage}
\begin{flushright}
TPI-MINN-92/69-T \\
BUTP-93/2
\end{flushright}
\vspace{0.2in}
\begin{center}
{\Large \bf Estimates of $m_d - m_u$ and $\langle\bar{d}d\rangle -
\langle\bar{u}u\rangle$ from QCD sum rules for $D$ and $D^{\ast}$ isospin
mass differences \\ }
\vspace{0.4in}
{\bf V.L. Eletsky$^{\dagger}$,} \\
Theoretical Physics Institute, University of Minnesota  \\
Minneapolis, MN 55455, USA \\
and \\
Institute for Theoretical Physics, Bern University \\
Sidlerstrasse 5, CH-3012 Bern, Switzerland \\
\vspace{0.2in}
{\bf and \\ }
\vspace{0.2in}
{\bf B.L. Ioffe \\ }
Institute of Theoretical and Experimental Physics \\
Moscow 117259, Russia \\
\vspace{0.5in}
{\bf   Abstract  \\ }
\end{center}
The recent experimental data on $D^{+}-D^{0}$ and
$D^{\ast\, +}-D^{\ast\,0}$ mass differences are used as inputs in the
QCD sum rules to obtain new estimates on the mass difference of light quarks
and  on the difference of their condensates:
$m_d -m_u =3\pm 1\, MeV$,
$\langle\bar{d}d\rangle -\langle\bar{u}u\rangle =
-(2.5\pm 1)\cdot 10^{-3}\langle\bar{u}u\rangle$ (at a standard
normalization point, $\mu = 0.5\, GeV$).

\vskip1.3in
\hrule height .2pt width 3in
\noindent$^{\dagger}$Permanent address: Institute of Theoretical and
Experimental Physics, Moscow 117259, Russia.
\end{titlepage}

The QCD sum rules invented more than a decade ago is now well known to
be a very useful tool to study properties of hadrons at
intermediate energies and to get
information on the basic parameters of QCD, such as quark masses and
non-perturbative condensates. One of the problems addressed already in the
pioneering papers\cite{svz} was the relation between the isotopic
symmetry violation on the level of hadrons and the difference
between $u$- and $d$-quark
masses and condensates. It was shown that the observed $\rho -\omega$
mixing implies $(m_d -m_u)/(m_u +m_d)\sim 0.3$ and
$\gamma\sim -1.5\cdot 10^{-2}$, where

\beq
\gamma =\langle\bar{d}d\rangle /\langle\bar{u}u\rangle -1
\label{gamma}
\eeq
ruling out a solution with
$m_u =0$, $m_d\neq 0$.
The quark mass difference was estimated before the advent of QCD sum
rules in refs.\cite{l,w} with the result
$m_d -m_u \approx 3\,MeV$ and
$m_d +m_u \approx 11\, MeV$.
The difference of condensates was later
estimated within the QCD sum rule framework in a number of papers\cite{sr}
with the result
$\gamma =-(3\, \div \, 10)\cdot 10^{-3}$.
The difference of condensates was also obtained  in the framework of
chiral perturbation theory\cite{gl}, which gives
$\gamma\approx -8\cdot 10^{-2}$, provided $m_d -m_u = 3\, MeV$,
$1-\langle\bar{s}s\rangle /\langle\bar{u}u\rangle\approx 0.2$ and
$m_s \approx 150\, MeV$.
Recently isospin violation in  QCD sum rules for the nucleon, $\Sigma$ and
$\Xi$ was considered\cite{di} and the following results were obtained:
$m_d -m_u = 3\pm 1 \, MeV$,
$\gamma =-(2\pm 1)\cdot 10^{-3}$.
Thus,
while most predictions for $m_d -m_u$ agree and are grouped around $3\, MeV$,
predictions for $\langle\bar{d}d\rangle -\langle\bar{u}u\rangle$ are more
diverse and range within an order of magnitude.
Moreover, arguments were given in ref.\cite{ch}
that $m_u$ may be equal to zero
due to instanton contributions to the renormalization of the quark mass.
Thus, additional independent estimates of differences between
$u-$ and $d-$quark masses and
condensates are certainly welcome.

In this paper we will consider QCD sum rules for
isospin mass splittings of $D^{\ast}$ and $D$ mesons and
make use of the recently reported\cite{ex}
new results on these splittings,

\begin{eqnarray}
m_{D^{\ast\, +}}-m_{D^{\ast\,0}}&=&3.32\pm 0.08\pm 0.05\,MeV   \nonumber\\
m_{D^{+}}-m_{D^{0}}&=&4.80\pm 0.10\pm 0.06\,MeV
\label{exp}
\end{eqnarray}
to obtain such estimates. The sum rules are similar to those which were
used in ref.\cite{be} to succesfully predict the mass splittings
$m_{D_{s}^{\ast}}-m_{D^{\ast}}=110\pm 20\,MeV$ and
$m_{D_s}-m_{D}=120\pm 20\,MeV$.

Let us start with the correlator of two pseudoscalar currents with quantum
numbers of $D$, $j_5 =\bar{c}\gamma_5 q$, where $q$ is either $u$, or $d$,
at Euclidean momentum $-q^2\gord 1\, GeV^2$,

\beq
C^{q}=i\int d^{4}x e^{iqx}
\langle 0|T\{j_{5}(x),j_{5}^{+}(0)|0\rangle
\label{cp}
\eeq
and consider its variation $\d C^q$ as the light quark mass rises from
zero to its actual value $m_q$.
To estimate $C^q$ it is
possible to take into account only the unit operator and the
quark condensate (Fig.1) in the operator product expansion, since the
contribution of operators of higher dimension to the heavy-light correlators
is negligible\cite{ae}. Using the expansion of the quark condensate
in the quark mass

\beq
\langle q_{\a}^{a}(x)\bar{q}_{\b}^{b}(0)\rangle =
\left( -\frac{1}{12}\d_{\a\b}
\langle\bar{q}q\rangle +\frac{i}{48}m_{q}\hat{x}_{\a\b}
\langle\bar{q}q\rangle\right)\d^{ab}
\label{qq}
\eeq
where $\langle\bar{q}q\rangle$ itself also depends on $m_q$,
it is easy to see that $\d C^q$ is a function of
$m_q$ and
$\langle\bar{q}q\rangle -\langle\bar{q}q\rangle _0$ where the subsript $0$
denotes the chiral limit.
On the other hand, saturating the correlators in the standard manner
by the corresponding lowest mass resonanses $D^{+}$ and $D^{0}$,
subtracting the continuum
from the contribution of the unit operator and applying the Borel
transformation\cite{svz}, $(s+Q^2)^{-1}\rightarrow M^{-2}\exp (-s/M^2)$,
we arrive at the following sum rule
for $\d C^d -\d C^u$

\begin{eqnarray}
\lefteqn{-\frac{\b_{D^{+}} - \b_{D^{0}}}{\b_{D}m_D} M^2+
(m_{D^+}-m_{D^0})_{hadr} = \frac{M^4 e^{m_{D}^2/M^2}}{2\b_{D}^2 m_D}\cdot }
\nonumber\\
& & \{ (m_d -m_u)[\frac{3m_c}{4\pi^2}
(e^{-x}-e^{-y}-x(E_{1}(x)-E_{1}(y)))L^{4/9}
-\frac{\langle\bar{q}q \rangle }{2M^2}e^{-x}
(1+\frac{m_c^2}{M^2})] \nonumber\\
& & +\frac{\langle\bar{d}d\rangle
-\langle\bar{u}u\rangle }{M^2} m_c e^{-x}L^{4/9}
-(s_{D^+} -s_{D^0})\frac{3}{8\pi^2 M^2 s_D}
(s_D -m_c^2)^2 e^{-y}\}
\label{ps}
\end{eqnarray}
Here $E_{1}(x)=\int_{x}^{\infty} dt\,t^{-1}e^{-t}$,
$m_c$ is the charmed quark mass, $s_D$ is the continuum threshold,
$x=m_c^2/M^2$, $y=s_D/M^2$
and the residue of $D$ meson into the current $j_5$ is
$\b_{D}^{2} = f_D^2 m_D^4/m_c^2$, where $f_D$ is the semileptonic
decay constant defined by
$\langle D|\bar{c}\gamma_{\mu}\gamma_5 q|0\rangle =-if_D p_{\mu}$. The
dependence of light quark masses and condensates on the normalization point
$\mu$ in the operator product expansion is given by powers of
$L=\ln (M/\Lambda)/\ln (\mu/\Lambda)$ where $\Lambda=150\, MeV$ and we take
$\mu =0.5\, GeV$ which corresponds to
$\langle\bar{q}q\rangle =-(0.24\, GeV)^3$.
Here $\b_{D^{+}}=\b_{D}+\d\b_{D}^d$, $\b_{D^{0}}=\b_{D}+\d\b_{D}^u$ and
$s_{D^{+}}=s_{D}+\d s_{D}^d$, $s_{D^{0}}=s_{D}+\d s_{D}^u$
where $\b_{D}$ and $s_{D}$ are the residue and the continuum
thresholds in the chiral limit and $\d\b_D^{u,d}$ and $\d s_D^{u,d}$
are the deviations of residues
and of continuum thresholds
from their values in the chiral limit.

Similar sum rules can be written for the correlator of two vector currents,
$j_{\mu} =\bar{c}\gamma_{\mu}q$,

\beq
C^{q}_{\mu\nu}=i\int d^{4}x e^{iqx}
\langle 0|T\{j_{\mu}(x),j_{\nu}^{+}(0)|0\rangle
\label{cv}
\eeq
In the case of the tensor structure $q_{\mu}q_{\nu}$ for which the sum rule
is known to work better\cite{ek}, we obtain

\begin{eqnarray}
\lefteqn{-(\frac{\b_{D^{\ast\, +}} -\b_{D^{\ast\, 0}}}
{\b_{D^{\ast}}m_{D^{\ast}}}-
\frac{(m_{D^{\ast\,+}}-m_{D^{\ast\,0}})_{hadr}}
{m_{D^{\ast}}^2}) M^2+
(m_{D^{\ast\,+}}-m_{D^{\ast\,0}})_{hadr}= } \nonumber\\
& & \mbox{} -\frac{m_{D^{\ast}}e^{m_{D^{\ast}}^2/M^2}}{2\b_{D^{\ast}}^2}
[(m_d -m_u)\langle\bar{q}q\rangle e^{-x}+
(s_{D^{\ast\, +}} -s_{D^{\ast\, 0}})
\frac{e^{-y}M^2}{4\pi^2}
(1-\frac{3m_c^4}{s_{D^{\ast}}^2}+\frac{2m_c^6}{s_{D^{\ast}}^3})]
\label{v}
\end{eqnarray}

It is important to emphasize that since we do not take into account
perturbative two-loop diagrams with one photon exchange in the
"theoretical" part of the sum rules,
only the hadronic parts of the isospin splittings, $(\Delta m_D)_{hadr}$ and
$(\Delta m_{D^{\ast}})_{hadr}$, enter eqs.(\ref{ps}) and (\ref{v}). To obtain
them we
use a quark model estimate\cite{leu} for the photon cloud part of the mass
difference $m_{D^{+}} -m_{D^0}= 1.7\pm 0.5\, MeV$ and also take into account
the electromagnetic hyperfine splitting

\beq
\delta m=-\frac{2\pi Q_c Q_q |\Psi (0)|^2}{3m_c m_q} [2S(S+1)-3]
\label{hf}
\eeq
where $S$ is the spin of the meson.
The quark masses here are the constituent ones, $m_c\sim 1.7\, GeV$,
$m_q\sim 0.3\, GeV$.  The quark-antiquark wave function at the origin may be
estimated using the relation
$f_D^2=12|\Psi (0)|^2/m_D$. Using the estimate\cite{ae} $f_D=170 MeV$,
we get that the hyperfine electromagnetic interaction contributes
$\sim 0.5\, MeV$ to $D^{+}-D^0$ and $\sim -0.17\, MeV$ to
$D^{\ast\,+}-D^{\ast\,0}$ mass splittings. Using the full experimental
mass differences from eq.(\ref{exp}) we thus have the mass splittings to be
used
in the sum rules in eqs.(\ref{ps}) and (\ref{v})

\begin{eqnarray}
(m_{D^{\ast\, +}}-m_{D^{\ast\,0}})_{hadr}&=&1.8\pm 0.5\,MeV   \nonumber\\
(m_{D^{+}}-m_{D^{0}})_{hadr}&=&2.6\pm 0.5\,MeV
\label{had}
\end{eqnarray}
{}From the sum rules in eqs.(\ref{ps}) and (\ref{v})
it follows that in the working region
for the Borel parameter $M^2$, the left hand sides should be close to linear
functions in $M^2$ whose extrapolations to $M^2=0$ give the corresponding
hadronic mass splittings while the slopes give information on the
residue differences, $\b_{D^{+}} - \b_{D^{0}}$ and
$\b_{D^{\ast\, +}} -\b_{D^{\ast\, 0}}$.
To numerically analize the sum rules we take
$m_D =1.87\, GeV$, $m_D^{\ast}=2.01\, GeV$,
the standard value of the quark
condensate, $\langle\bar{q}q\rangle =-(0.24)^3\,GeV^3$, and
$m_c =1.35\, GeV$. For the residues and continuum thresholds we take the
estimates obtained with sum rules in refs.\cite{ae,ek},
$\b_D^2\approx\b_{D^{\ast}}^2\approx 0.2\, GeV^4$ and
$s_D\approx s_{D^{\ast}}\approx 6\, GeV^4$.

Thus, $m_d -m_u$, $\langle\bar{d}d\rangle -\langle\bar{u}u\rangle$, the
Borel parameter $M^2$ and the differences of continuum thresholds,
$s_{D^+} -s_{D^0}$ and $s_{D^{\ast\, +}} -s_{D^{\ast\, 0}}$
are the fitting parameters of the sum rules in eqs.(\ref{ps}) and (\ref{v}).
The working regions in $M^2$ determined by the requirement of
controllable contributions from non-perturbative power corrections and
continuum in the chiral limit for $u$- and $d$-quarks are\cite{ae,ek}
$1\, GeV^2\lord M^2\lord 2\, GeV^2$ for the pseudoscalar channel and
$1.5\, GeV^2\lord M^2\lord 2.5\, GeV^2$ for the vector channel.
We will let the difference of thresholds vary around an estimate
$s_{D^{+}} -s_{D^{0}}\sim s_{D^{\ast\, +}} -s_{D^{\ast\, 0}}
\sim (m_d -m_u)\sqrt{s_D}$ which implies that the threshold and
the quark mass differences are of the same sign.

In Fig.2 we show the result of the sum rule calculation of the $D^{\ast}$
mass difference by plotting
$f_{V}(M^2)-M^2 df_{V}(M^2)/dM^2$,
where $f_{V}(M^2)$ is the r.h.s. of eq.(\ref{v}).
{}From eq.(\ref{v}) one sees that only the quark mass difference enters this
sum rule and one might hope to fix $m_d -m_u$ from it.
The term proportional to $s_{D^{\ast\, +}} -s_{D^{\ast\, 0}}$
has the same sign as the term proportional to $m_d -m_u$ and in the
whole working interval in $M^2$ contributes less than $50\%$
to $f(M^2)-M^2 df(M^2)/dM^2$ at $m_d -m_u = 3\, MeV$.
The values $m_d -m_u \gord 4\, MeV$ cannot be made compatible with
$(m_{D^{\ast\, +}}-m_{D^{\ast\,0}})_{hadr}$ from eq.(\ref{had})
at any $s_{D^{\ast\, +}} -s_{D^{\ast\, 0}}>0$.
At $m_d -m_u \lord 4\, MeV$ the compatibility can be achieved by
varying $s_{D^{\ast\, +}} -s_{D^{\ast\, 0}}$. However, at
$m_d -m_u \lord 2\, MeV$ the continuum term is dominant in
the whole working region in $M^2$ and the sum rule is not reliable.

As for the $D$ mass difference,
the corresponding sum rule in eq.(\ref{ps}) is contributed
both by $m_d-m_u$ and $\langle\bar{d}d\rangle -\langle\bar{u}u\rangle$.
In Fig.3 we show the numerical results for
$m_d -m_u = 3\, MeV$ and two different values of
$\gamma =\langle\bar{d}d\rangle /\langle\bar{u}u\rangle -1$.
One can see that the value $\gamma = -2.5\cdot 10^{-3}$ is consistent
with the $(m_{D^{+}}-m_{D^{0}})_{hadr}$ from eq.(\ref{had}),
while a higher value
$\gamma =-6\cdot 10^{-3}$ is definitely excluded.

In summary, we conclude from our numerical analysis
of the sum rules for isospin mass splittings
in $D$ and $D^{\ast}$ mesons that $m_d -m_u = 3\pm 1 \, MeV$
and $\gamma = -(2.5\pm 1)\cdot 10^{-3}$.
The result for $m_d -m_u$ is consistent with the earlier estimates
and together with the relation
$(m_u +m_d)\langle\bar{u}u\rangle =-F_{\pi}^{2}m_{\pi}^{2}$
excludes the option $m_u=0$ advocated in ref.\cite{ch}.
Our estimate for the condensate difference parameter
$\gamma$ supports the value obtained in ref.\cite{di}
from the isospin splittings in baryons and
differs from results of earlier papers\cite{sr,gl}\footnote
{For a discussion of possible reasons for this disagreement,
see ref.\cite{di}.}.

We are grateful to H. Leutwyler for useful discussions. One of the
authors (V.L.E.) acknowledges the warm hospitality extended to him
at the Universities of Minnesota and Bern.
This work was
supported in part by the US Department of Energy under contract
DE-FG02-87ER40328 and by Schweizerischer Nationalfonds.

\pagebreak
{\large\bf Figure Captions:}
\begin{itemize}
\item
Fig. 1: The diagrams taken into account in the sum rules.
\item
Fig. 2: Sum rule for $D^{\ast}$ mass splitting:
        $f_{V}(M^2)-M^2 df_{V}(M^2)/dM^2$ versus $M^2$,
        where $f_{V}(M^2)$ is the r.h.s. of eq.(\ref{v})
        for $m_d -m_u = 1\, MeV\; (a)$, $3\, MeV\; (b)$ and
        $5\, MeV\; (c)$.
        In each of the three cases the lower and the upper curves
        correspond to $s_{D^{\ast +}}- s_{D^{\ast 0}}= 0$ and
        $0.005\, GeV^2$, respectively. The dotted horizontal lines
        are the boundaries set by eq.(\ref{had}).
\item
Fig. 3: Sum rule for $D$ mass splitting:
        $f_{P}(M^2)-M^2 df_{P}(M^2)/dM^2$ versus $M^2$,
        where $f_{P}(M^2)$ is the r.h.s. of eq.(\ref{ps})
        for $m_d -m_u = 3\, MeV$  and a) $\gamma = -2.5\cdot 10^{-3}$;
        b) $\gamma = -6\cdot 10^{-3}$. The numbers at the curves correspond
        to the value of $s_{D^{+}}- s_{D^{0}}$ (in $GeV^2$).
        The dotted horizontal lines
        are the boundaries set by eq.(\ref{had}).
\end{itemize}
\end{document}